\documentclass{emulateapj}

\usepackage{natbib}
  \bibpunct{(}{)}{;}{a}{}{,} 

\shorttitle{ASTMON}
\shortauthors{Aceituno et al.}

\begin{document}

\title{An All Sky Transmission Monitor: ASTMON}

\author{J.~Aceituno\altaffilmark{1},
S.F~S\'anchez\altaffilmark{1}, F. J. Aceituno\altaffilmark{2},
D.~Galad\'i-Enr\'iquez\altaffilmark{1},
J.J.~Negro\altaffilmark{3}, R.C.~Soriguer\altaffilmark{3}, G.
Sanchez Gomez\altaffilmark{4} }
 \altaffiltext{1}{Centro Astron\'o mico Hispano-Alem\'an (CAHA A.I.E), Jes\'us Durb\'an Rem\'on
2-2, 04004 Almer\'\i a, Spain} \altaffiltext{2}{Instituto de
Astrof\'\i sica de Andaluc\'\i a (IAA), Consejo   Superior de
Investigaciones Cient\'\i ficas (CSIC), C/ Camino Bajo de Hu\'etor
50, E-18008, Granada, Spain} \altaffiltext{3}{Estaci\'on
Biol\'ogica de Do\~nana, CSIC}
\altaffiltext{4}{SATEC}\email{aceitun@caha.es}

\begin{abstract}

We present here the All Sky Transmission MONitor (ASTMON),
designed to perform a continuous monitoring of the surface
brightness of the complete night-sky in several bands. The data
acquired are used to derive, in addition, a subsequent map of the
multiband atmospheric extinction at any location in the sky, and a
map of the cloud coverage. The instrument has been manufactured to
afford extreme weather conditions, and remain operative. Designed
to be fully robotic, it is ideal to be installed outdoors, as a
permanent monitoring station. The preliminary results based on two
of the currently operative units (at Do\~nana National Park -
Huelva- and at the Calar Alto Observatory - Almer\'\i a -, in
Spain), are presented here. The parameters derived using ASTMON
are in good agreement with previously reported ones, what
illustrates the validity of the design and the accuracy of the
manufacturing.  The information provided by this instrument will
be presented in forthcoming articles, once we have accumulated a
statistically amount of data.

\end{abstract}


\section{Introduction}

The night sky brightness, the number of clear nights, the seeing,
transparency and photometric stability are some of the most
important parameters that qualify a site for front-line
ground-based astronomy (Taylor et al. 2004). There is limited
control over all these parameters, and only in the case of the sky
brightness is it possible to keep it at its natural level by
preventing light pollution from the immediate vicinity of the
observatory (Garstang 1989). Previous to the installation of any
observatory, extensive tests of these parameters are carried out
in order to find the best locations, maximizing then the
efficiency of these expensive infrastructures (Thomas-Osip et al.
2010, Schock et al. 2009) . However, most of these parameters are
not constant along the time, neither in short term nor in long
term time scales.

An on-line monitoring of all these conditions has been proved to
be a fundamental tool to take decisions on the observational
strategies, in order to increase the efficiency of any
professional observatory (Travouillon et al. 2011). In particular,
a very interesting input to decide about the optimal strategy
would be to estimate in real-time the stability of the atmosphere
and the location of the darkest and the most transparent areas in
the sky (eg, Benn \& Ellison et al. 1998a,b; S\'anchez et al.
2007; Moles et al. 2010; High et al. 2010; Zou et al. 2010).

Along the 26 years of operations of the Calar Alto observatory,
there has been different attempts to characterize some of the main
properties described before: (i) Leinert et al. (1995) determined
the sky brightness corresponding to the year 1990; (ii) Hopp \&
Fernandez (2002) studied the extinction curve corresponding to the
years 1986-2000; However, we lack of a consistent study of all of
them, spanning over a similar time period.This information becomes
even more important for long-term surveys, where automatic or
semi-automatic schedulers are programmed in order to optimize the
observation over large areas in the sky (eg., Moles et al. 2008).

In order to characterize these parameters most of the professional
observatories have installed permanent monitors which provide an
almost real-time estimation of them. The values produced by these
monitors are frequently accessible on-line via web services,
stored in databases, and even included in the headers of the
images obtained at the observatory. They compose an extremely
useful database on the long term stability of the night-sky
conditions at a certain site. In general, most of these monitors
produce global parameters, based on large aperture measurements,
like the cloud monitors based on the temperature difference
between the ground layer and the sky, or local parameters, like
the information provided by most of the seeing and transparency
monitors: eg., RoboDIMM (Aceituno 2004) or EXCALIBUR (S\'anchez et
al. 2007, Moles et al.  2010). On other occasions, all sky
monitors provide qualitative estimations on the sky transparency,
based on direct images to estimate the cloud coverage, but no
quantitative ones.

As part of a large program to determine the observing conditions
at different locations in Spain (S\'anchez et al. 2007, 2008;
Moles et al. 2010), we developed a new set of instruments, able to
estimate the most important parameters that determine the quality
of an astronomical site, working in fully automatic mode. The
first of these instruments was named EXCALIBUR (P\'erez-R\'amirez et  al. 2008), a multi-wavelength
extinction monitor, whose main purpose was to determine the
atmospheric attenuation curve, and derive the relative
contribution of each its components, at a particular location in
the sky. The main idea behind this instrument was to determine
these parameters at the sample position where the astronomical
observation is taking place, monitoring any possible change in the
night-sky stability.

This instrument provides accurate local information, but it cannot
provide simultaneous information on the considered parameters at
different locations in the sky. For doing so, a different concept
of instrument is required. In this article we present the All Sky
Transmission MONitor (ASTMON), an instrument that, working on the
same principals of EXCALIBUR, derives the surface sky brightness,
atmospheric extinction and extinction curve at several locations
in the sky simultaneously. It also provides additional information
about the night-sky quality, based on the measurements acquired,
like the photometric conditions, cloud-coverage and amount of
light-pollution.

Although the main goal of these instruments is to determine the
night-sky quality for the astronomical observation, this
information has been found useful in another science fields. Due
to that, a prototype of a EXCALIBUR unit was installed in the
Andalusian Center of Environment Studies (Granada), to estimate
the light pollution and aerosol content in populated areas
(P\'erez-Ram\'\i rez et al. 2008). In addition, the first
prototype of ASTMON was installed permanently in the Do\~nana
Biological Reserve, CSIC (Spain), to determine the effects of the
light pollution in the biological environment of this protected
area. Both instruments were calibrated at the Calar Alto
observatory, which has acquired and installed units of both of
them, integrating them in their monitor systems. The calibration
runs demonstrated that these instruments are able to provide
quantitative valuable measurements in a completely automatic way,
comparable to the values provided by other monitors and/or guided
astronomical observations.

The structure of this article is as follows: in Section \ref{con}
we present the basic concepts in which the instrument is based; in
Section \ref{inst} the instrument itself is described,
illustrating its main components; in Section \ref{data} we present
the results from the calibration run, comparing them with those
derived by other well tested monitors and/or published data. In
Section \ref{sum}, we present a summary of the results from the
experiment and the future perspective of this kind of instruments.

\section{Basic concepts of the instrument}
\label{con}

The procedures adopted to derive the extinction and night-sky
surface brightness for this instrument are based on simple
photometric concepts (Falchi 2010). The first step consists of the
acquisition of a full sky frame at a particular optical band, by
the techniques that will be described below. The exposure time
should be large enough to sample the sky-brightness in the areas
free of objects with enough signal-to-noise ratio to derive a
reliable measurement, and, at the same time, short enough not to
saturate a sufficient number of bright stars, that are used as
photometric self-calibrators.

A source detection algorithm is then run over the considered
image, providing a catalogue of the stars detected in the field.
Adopting an a-priori calibrated distortion solution, the catalogue
of detected stars is cross-matched with an astrometric and
photometric calibrated catalogue of stars. This procedure allows
both to identify the stars and to estimate the corresponding
airmass within the considered image. Then, the flux corresponding
to each of these stars is derived by adopting a simple aperture
photometry algorithm on the image. The contribution from the sky
emission is estimated by measuring the mode of the counts within a
predefined annular ring around each star, correcting for the
aperture differences, and subtracted to the flux measured in the
central aperture. Different experiments have shown us that the
mode is the most optimal statistical estimator of the surface
brightness of the sky-background, when dealing with crowded fields
and large pixel scales.
  \begin{figure}
\resizebox{\hsize}{!}
{\includegraphics[width=\hsize]{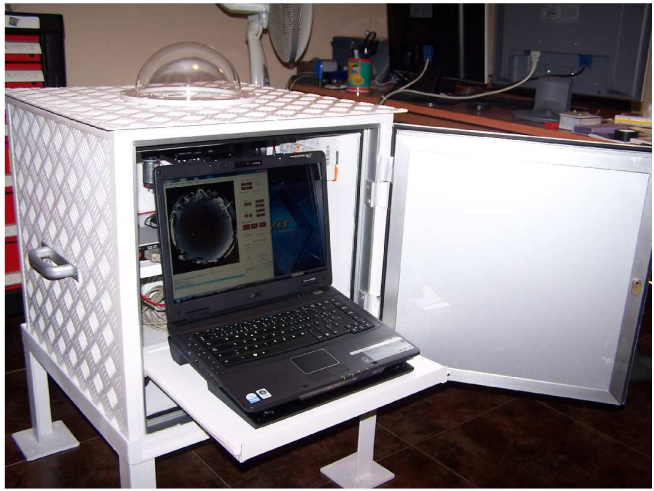}}
  \caption{\label{astmon.}
 The figure shows an external view of an ASTMON unit. The internal components
 are protected by a weather - proof housing, so the instrument can be
 installed out-doors.
 }
  \end{figure}

The described procedure provides us with a catalogue of detected
stars. For each of them, the catalogue includes the coordinates of
the star projected on the sky, its airmass, its calibrated
photometry at the considered band, and the observed flux (in
counts). These catalogues can be used to derive the instrumental
zero-point ($ZP$) and the extinction coefficient for the
considered band ($\kappa$), applying the following classical
formulae:

\begin{equation}
 2.5  {\rm log}_{10} (F/t) + mag = ZP - \kappa * \chi
\end{equation}

where $F$ is the number of counts derived for the central
aperture, $t$ is the exposure time, $mag$ is the catalogued
magnitude for the considered star, and $\chi$ is the air-mass of
the observed star. The extinction, that depends on the particular
atmospheric conditions, is highly variable. However, the
zero-point, that depends mostly on the characteristics of the
instrument, is more stable with time. Under photometric
conditions, in which the extinction coefficient is uniform at
almost any location in the sky, it is possible to derive both
parameters by performing a classical linear square-regression
between the instrumental magnitude ($2.5 log (F/t) + mag $) and
the airmass ($\chi$).  The wide airmass range covered by the
instrument, allows to obtain an accurate estimation of both
parameters with a single image. Although night-length photometric
conditions are scarce, it is relatively easy to find short periods
of photometric conditions, and therefore, it is possible to obtain
several estimations of the zero-point in any clear night.

\begin{figure}
\resizebox{\hsize}{!}
{\includegraphics[width=\hsize]{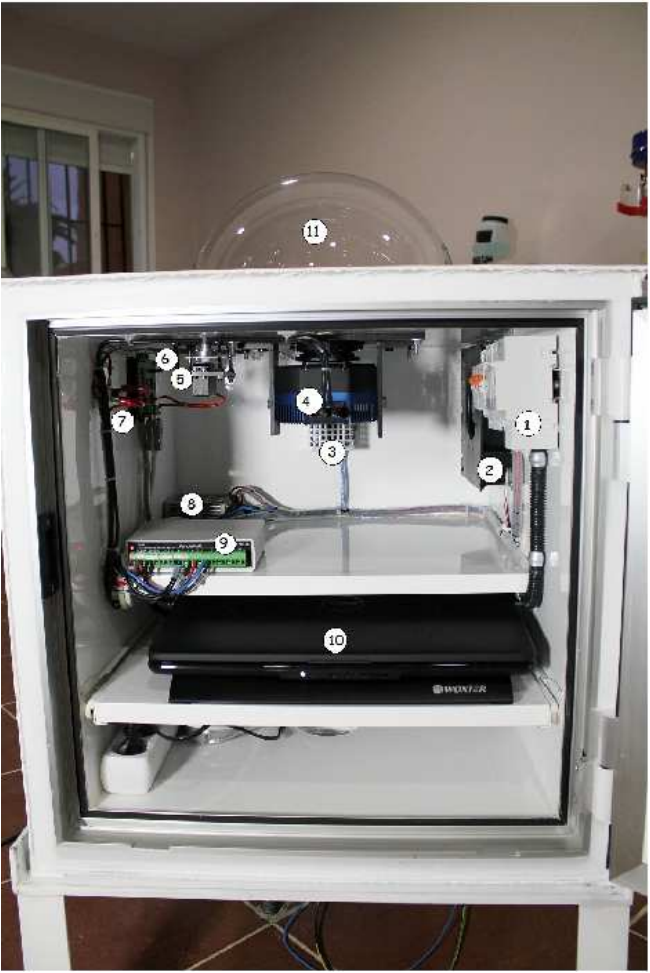}}
  \caption{\label{electronic}
The figure shows the internal view of the system: (1) Ground fault
and electrical breakers. (2) Main fan. (3) Anti-humidity system.
(4) CCD head with filter wheel. (5) Servo-motor of the solar
shutter. (6) Servomotor controller. (7) Thermostats. (8) Main
stabilized power supply. (9) PIC I/O controller. (10) Computer.
(11) Acrilic dome.
 }
  \end{figure}

Once obtained the zero-point, the extinction is derived at each
location in the sky, by solving the previous equation for
$\kappa$. In addition to that, for each area free of targets it is
possible to derive the sky-surface brightness, by applying the
formulae:

\begin{equation}
 SB = ZP - 2.5 {\rm log}_{10}  (F_{sky}/A)
\end{equation}

where $SB$ is the night-sky surface brightness in mag/arcsec$^2$,
$ZP$ is the zero-point at the considered band and $F_{sky}$ is the
number of counts per second measured in a certain area $A$ (in
square arcsec). This later parameter is the only one that presents
a certain difficulty, since in all-sky monitors the area covered
by each pixel is not uniform, and therefore an accurate area
correction has to be applied, based on the estimation of the field
distortion.

\section{Description of the Instrument}
\label{inst}

\subsection{The hardware}

The technical design of the instrument is rather simple. It
comprises (1) a fisheye to sample the full sky ($\sim$180$^o$) in
a single shot; (2) a filter-wheel, where the optical-band filters
are located; (3) a CCD detector, including a shutter; (4) a
electronics system that controls the different behaviors of the
instrument and (5) a control and storage computer (a standard
Core-Duo based laptop). All the elements are packed in a enclosure
prepared to afford extreme weather conditions.

  \begin{figure}
\resizebox{\hsize}{!}
{\includegraphics[scale=0.3]{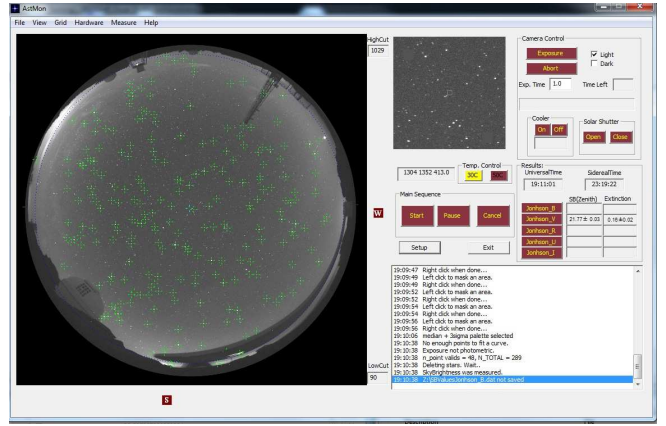}}
  \caption{\label{photometric1}
Snap-shot of the custom software developed for ASTMON, taken
during one of the calibration runs. Different control buttons,
entries, and panels showing the status of the instrument are seen
at the right-side of the program. The panel at the left side shows
one particular image of the night-sky. The green circles indicate
the location of the detected stars that match with the
corresponding ones at the astrometric/photometric catalogue, as
described in the text.
 }
  \end{figure}

Figure 1 shows an out-side view of the instrument, at the Calar
Alto labs. The white enclosure has its front door open, and the
control computer is seen, and the glass closing window on top of
the light entrance. A solar shutter is placed just above the
fisheye lens to protect it against the direct exposure to the
sunshine. As indicated before, the enclosure is designed to afford
extreme weather conditions, providing an almost isolated
environment when it is closed. It is equipped with double isolated
walls, and a drying system based on Chloride Calcium that keeps
the humidity inside at about of 50-60\% (ie., the optimal one to
operate the electronics and CCD). The design maintains the
instrument operational under high relative external humidities
(near 100\%), and prevents from any damage  the internal cabling
and components.

The imaging system consists of a fisheye lens that provides a
field of view of 180 degrees, covering the full sky, and a camera.
The camera mounts a KODAK KAF-8300 CCD of 8.3 megapixels, with
tiny 5.4 microns pixels (figure 2). This results in a pixel scale
of 3.8 arcminutes per pixel.  A filter-wheel, with 5 different
holders, is located between the fisheye lens and the camera. A set
of standard Johnson filters is used, included U, B, V, R and I for
the Calar Alto unit and B, V and I for the Do\~nana unit. However
any set of filters that matched the holders (1 1/4 inches
diameter), can be installed if required.

The temperature within the enclosure is sampled by four sensors,
installed at different locations. If the temperature rises to a
critical value (due to the combined heating of the CCD,
electronics and computer), and automatic fan activates to extract
the hot air through a pipe designed for this purpose. This pipe
includes a system to avoid that rain or snow gets into the
enclosure. On the other hand, if the temperature drops below a
critical value, a heating system is activated. The whole system
forces the temperature to be stable within a range of optimal
values, under any external conditions.  In addition to that, an
humidity sensor monitors the value of the current internal
humidity, and disconnects the system if a critical value is
reached.

The main electronics comprise an USB relay board and a set of
digital/analog inputs based on a micro-pic controller. This system
allows the computer to take control on the internal power of all
the devices. In this way it is easy to deactivate the system
during day time and to reactivate it again when the sun is set.
The electronics comprises too a servo motor controller for the
solar shutter and 4 analog inputs to read the values of the
temperature sensors described before. Finally, a ground fault and
electrical breakers have been included for security.

Figure 2 shows the interior of the instrument, from the main door
towards the inside of the enclosure. The different elements
described before have been labelled, illustrating their location
within the system.

 \begin{figure}
\resizebox{\hsize}{!}
{\includegraphics[width=\hsize]{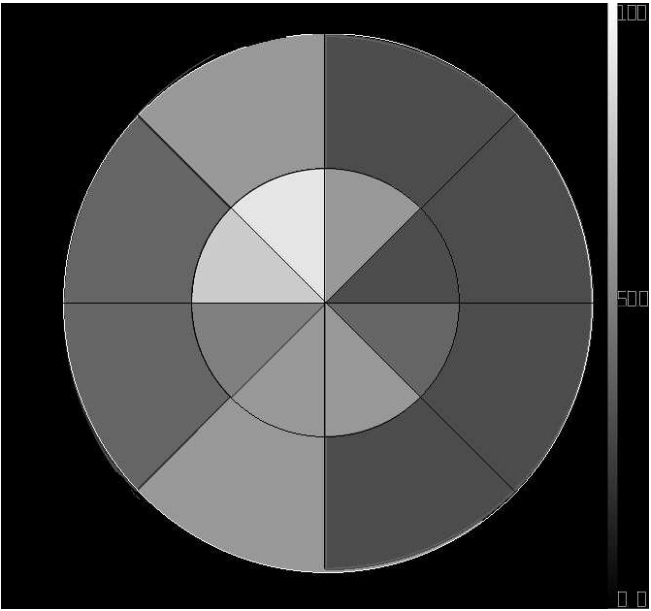}}
  \caption{\label{CM}
Map of the clouds coverage generated by ASTMON during a middly
cloudy night at the Do\~nana National Park. The greycolor scale
represents the percentage of cloud's coverage, where the darker
regions correspond to the clearer sky areas.
 }
  \end{figure}

\subsection{The software}
\subsubsection{Control system and starting up}

The electronics of the system are controlled by the computer
through a custom designed software, which allows the instrument to
be completely robotic.  The software switches on the system on the
basis of the computers time and date, by checking if the
astronomical night has started. In a similar way, it switches off
the system, when the astronomical night has gone. To switch on the
system, the computer activates all the powers and enables the
communication between the CCD and the filter wheel. Then, it
activates the cooling system of the CCD, in order to reach the
optimal operational temperature of $-$15$^o$ Celsius. Once this
temperature is reached, the program generates a master bias by
taking a set bias frames, and getting the median. Dark current is
particularly high for this particular CCDs. In order to remove
this from the science frames, a set of dark frames with different
integration times is derived, and stored, during the startup
procedure. In this way, it is not needed to take a dark frame for
each science observation.

  \begin{figure}
\resizebox{\hsize}{!}
{\includegraphics[scale=0.3]{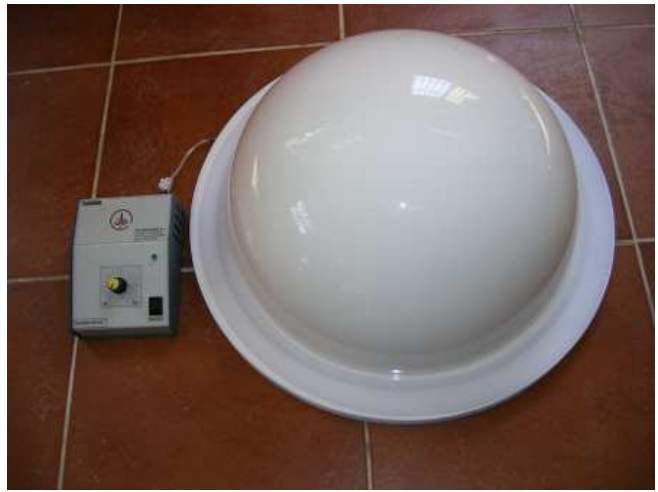}}
  \caption{\label{photometric2}
Picture of DomeLight. This device has been developed to shine
uniformly any wide-field camera.
 }
  \end{figure}

Once performed the startup procedure, the program starts the
acquisition of data, following a predefined sequence. Each step in
the sequence comprises the selection of a certain filter, the
acquisition of a set of images, and their reduction and analysis.
The only parameter to be adjusted of these sequence is the
exposure time of the frames taken with each filter. The final
selected exposure times are 300 seconds for the U-band and 40
seconds for remaining ones (B,V,R and I). These values were
derived experimentally, driven by the compromise between taking
exposures short enough not to be affected by the rotation of the
Earth (which affects to the astrometric solution and the shape of
the stars on the images), and getting, on the other hand, enough
signal-to-noise ration in both the stars and the night-sky
background. The data are corrected for the bias and master dark
immediately after acquiring them. After that, a master flatfield
correction is applied (we will describe later how it is obtained).
Finally, a mask is applied to remove from the images any artifact
due to local buildings, trees, and so on. This mask is generated
manually, by a routine included in the software. This procedure
has to be done only once, when the instrument is operated for the
first time at a particular location.

\subsubsection{Astrometric solution. }

Once a particular frame is obtained for a particular band, and the
data are reduced, the program starts to analyze it. First, it
performs an automatic identifications of the stars in the image.
The program uses the Ducati (2002) photometric catalog, that
comprises of accurate astrometric positions and photometry for
most of the stars down to 6th magnitude in all the Johnson bands.
However, any other catalogue can be selected by the user, by
simply marking it in the program interface.

Fisheye lenses suffer from strong geometric optical aberrations,
that have to be corrected prior to finding the astrometric
solution. The correction required for the geometric aberration is
derived empirically, being required to be obtained only once,
during the calibration runs of the the instrument. For deriving
this calibration the software includes a routine that allows to
display on top of the acquired night-sky image the theoretical
positions of a hundred of stars in the field of view. Then, the
user is requested to match the estimated and real positions of the
stars within the field, by clicking on them. The coordinates of
this manual selection are used as a first approximation to the
final ones. Those are derived on the basis of an automatic
centroid algorithm, which looks for the brightest star within a
diameter of 8 pixels (30 arcminutes) with respect to the manually
selected coordinates. When more than one bright star is found,
they are discarded from the final catalogue. Fortunately the
probability of getting several stars brighter than 6th magnitudes
in a region of 30 arcminutes is low. The adopted centroid
procedure has an accuracy of $\sim$1 pixel (3.75 arcminutes),
sufficient for its purpose.

After this matching process, the software evaluates the radial
optical aberration of the fisheye by fitting a second order
polynomial function to the differences between the estimated and
observed positions with respect to the radial distance. This
solution is stored to be applied as part of the astrometric
solution to the images.

  \begin{figure}
\resizebox{\hsize}{!}
{\includegraphics[width=\hsize]{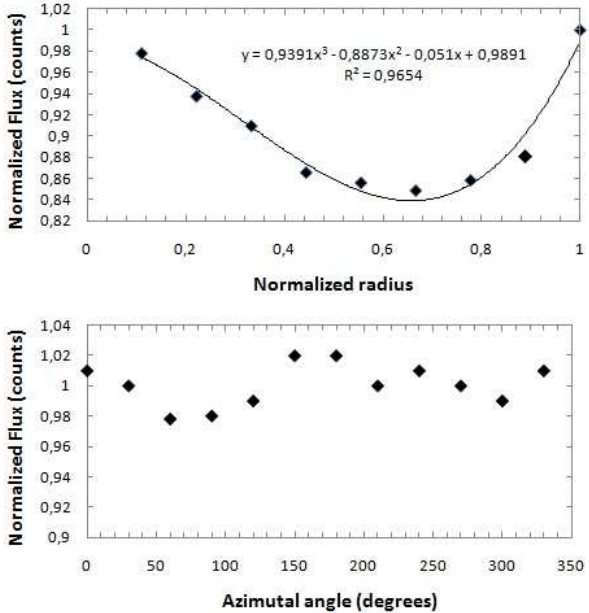}}
\caption{\label{photometric3}
Estimation of the illumination correction of the DomeLight device.
The top panel shows the radial distribution of the intensity
measured with the two apertures sensor, and the bottom panel shows
the azimutal one. In both cases the intensity is normalized to the
intensity at the zenith. The solid line in the left panel shows
the best fitted 2nd order polynomial function.
}
  \end{figure}

\subsubsection{Diagram of the coverage by clouds}

As a result of the plate solution, a cloud map is generated for
each acquired frame. To derive this cloud map, the sky image is
divided in 16 areas, following the pattern shown in Figure 4. The
number of detected stars is compared with the number of expected
(on the basis of the considered catalogue), within each of these
areas. The ratio between both parameters is used as an estimation
of the cloud coverage, assuming that stars are not detected due
mostly to the presence of clouds. The program represents the final
cloud-coverage diagram as a grayscale image, where the darker
sectors represent the areas of larger percentage of sky without
clouds (Figure 4). Altogether the procedure provides a method to
quantify the clouds's coverage.

\subsubsection{Map of the night sky surface brightness}

A classical aperture photometry is applied to each star detected
in the field, by measuring the counts within a circle with a
radius of 3 pixels, and two adjacent rings enclosed by circles
with a radius of 4 and 6 pixels, respectively.  The actual size of
these radii can be modified by the user. According to Equation 1,
the estimated extinction for all the stars is derived. Previously,
the instrumental Zero Point has been calculated during a
calibration run.

   \begin{figure}
\resizebox{\hsize}{!}
{\includegraphics[width=\hsize]{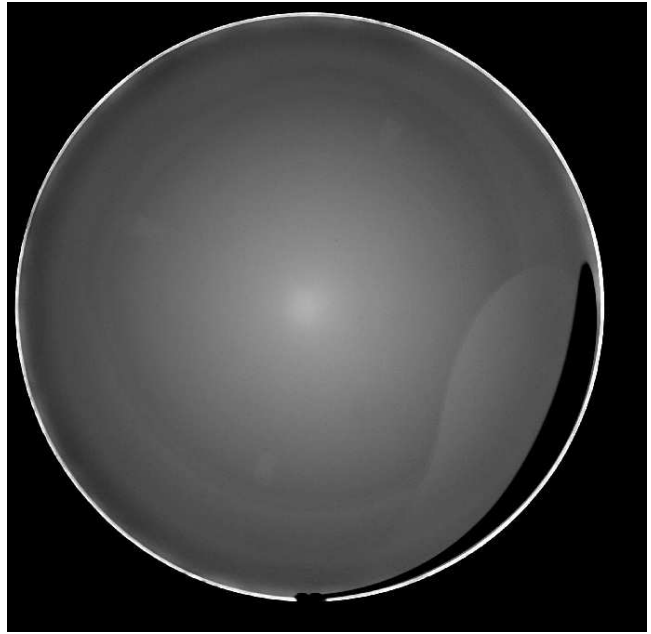} }
  \caption{\label{photometric4}
Flat field frame obtained with DomeLight in the V- band. The
effect of the solar shutter is shown in the image. This effect
also appears in the sky frames but it is corrected by the flat
fields.
 }
  \end{figure}

Afterwards, the stars are removed from the images by replacing the
values of the intensity of the corresponding pixels by the mode
intensity within the surrounding ones (Parker 1991). This cleaning
procedure is adopted to avoid the contamination due to these
sources. Finally, the night-sky brightness is derived for each
pixel on the basis of equation 2. The final frame is stored as
both FITS and JPEG files, which will allow the user to perform
both further analysis and/or display the results in webpages or to
produce movies with the provided information.

The software allows the user to define a set of positions in the
sky, in order to monitor the variations of the night-sky
brightness along the time at a particular location in the sky. The
size of the sampled region around each position is defined by the
user. Once selected, the program derives the mode of the night-sky
surface brightness within this region, and stores it in a
pre-defined ASCII file, for each acquired frame. The universal
time is also stored in this file, to facilitate any possible
long-term study of the light pollution at different locations in
the sky.

\begin{figure*}
\resizebox{\hsize}{!}
{
\includegraphics[width=\hsize,angle=0]{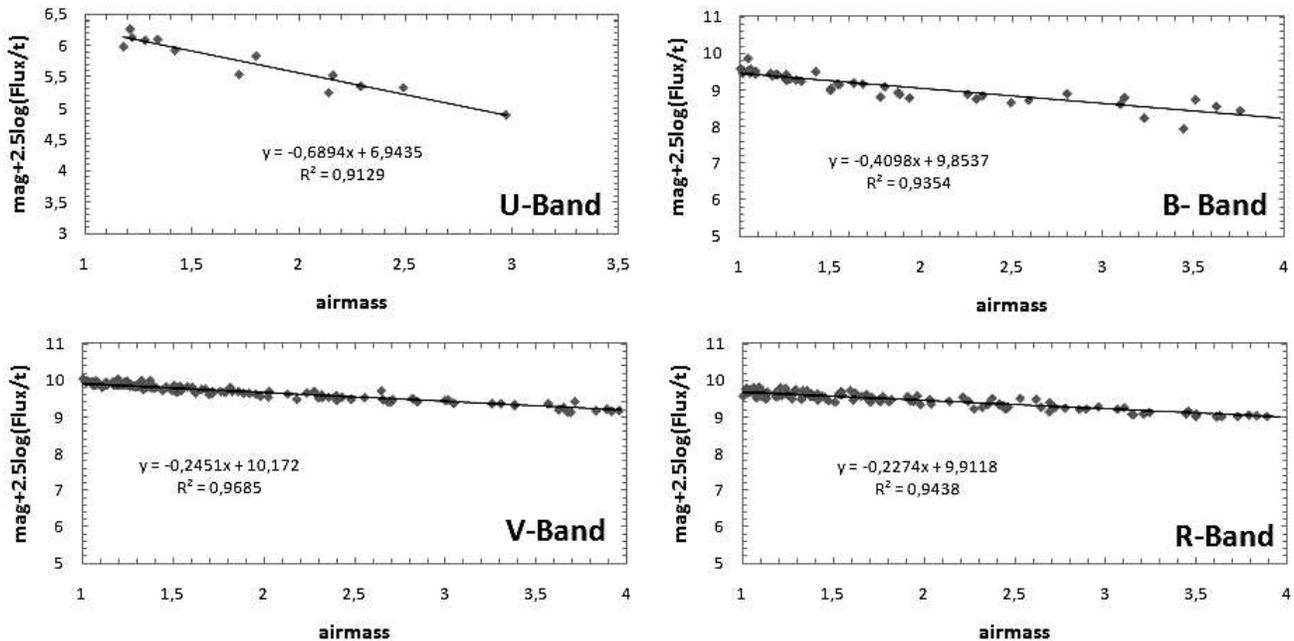}
}
 \caption{\label{sb}
Example of the linear regression technique used to derive both the
zero-point and the average extinction at a certain time for a
considered filter. Each panel shows the distribution of the
observed apparent magnitude versus the corresponding airmass for
four of the five filters installed in the instrument: U, B, V and
R-band. The data corresponds to the first calibration run with the
instrument at the Calar Alto observatory. The number of stars
detected per filter is considerably different, being 15, 57, 133
and 125, respectively. In this first calibration run only the
stars brighter than 4 magnitudes in the $V$-band were used.
 }
  \end{figure*}

\section{Results from the Calibration Run}
\label{data}

A calibration run was performed during the photometric night of
November 1st 2009, at the Calar Alto Observatory. The main goal of
this calibration run was to obtain the finally required master
calibration frames, derive the instrumental zero-points for the
different bands and gather preliminary data to compare the results
with previously reported ones.

\subsection{A new approach to get flat fields}

To obtain accurate flat field frames with a fisheye device is
difficult due to the complexity of illuminating 180 degrees with
an uniform light source. We designed an innovative procedure to
obtain such flat field frames based on a uniform whitened dome,
internally shined by a ring of white leds, located around the
dome, at its bottom. A white ring located on top of the leds
scatters the light within the dome, and produces the uniform
illumination. A picture of this device, so-called DomeLight, is
shown in Figure 5.

\begin{table}
\begin{center}
\caption{Zero points for the U, B, V, and I Johnson bands} \label{tab_mag2}
\begin{tabular}{ll}\hline
\tableline \tableline
U & 6.87 $\pm$0.15\\
B & 9.889$\pm$0.025 \\
V & 10.164$\pm$0.011\\
R & 9.906$\pm$0.015 \\
I & 8.06$\pm$0.21 \\
\tableline
\end{tabular}
\end{center}
\end{table}

The uniformity of the inner illumination can be checked with the
help of a CCD equipped with two small apertures faced up, near one
to each other. This device allows us to sample small areas of the
dome, while the contamination of the scattered light is minimized.
The flux is measured at different locations within the dome with
the help of this device. No dependency is found along the
different azimuthal angles, within a dispersion of just $\sim$5\%.
This dispersion corresponds to an error of less than 0.005
magnitudes in the derivation of the night-sky brightness, what has
been considered in the error budget. On the other hand, it is seen
a clear radial dependency of the illumination, that can be
modelled by a simple second order polynomial function. This
illumination correction is applied in the derivation of the flat
field obtained with this device.

Figure 6 shows the radial and azimuthal distributions of the illumination, as
derived on the basis of the procedure described before. An example of the
typical flat-field frame obtained prior to the illumination correction is shown Figure 7.

Finally, the flatfield is derived by placing the DomeLight device
on top of the protection transparent dome of the instrument. The
sizes of both domes are calculated in such a way that they match
each other. A set of exposures is taken with the lights of the
DomeLight switched on. After applying the illumination correction
a high quality master flat-field is derived. This master
flat-field is stored to be used in any further reduction analysis.

  \begin{figure}
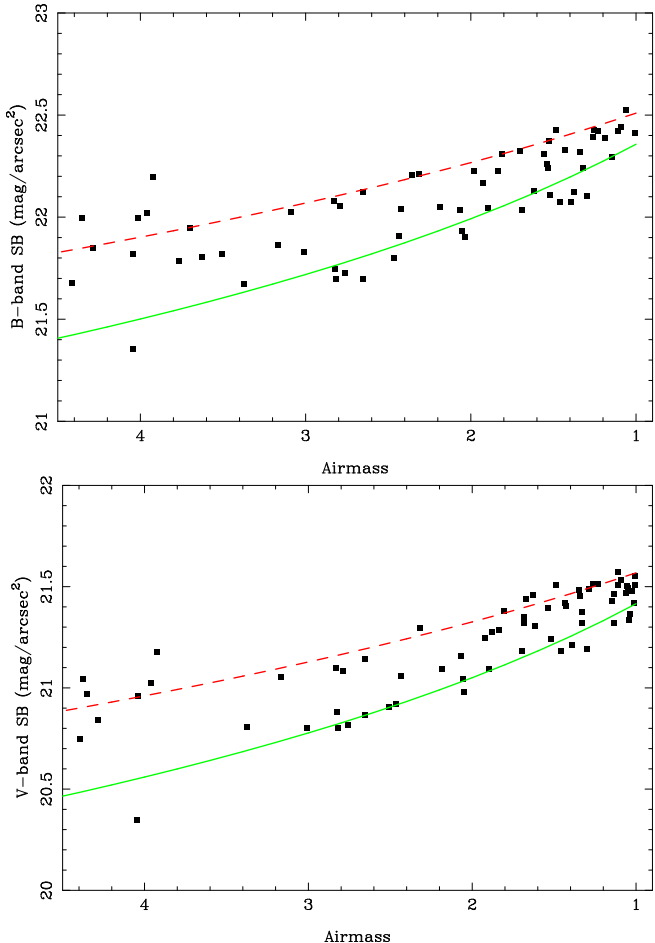

\resizebox{\hsize}{!}
{\includegraphics[width=\hsize,angle=270]{aceitunoFig9.ps}}
\resizebox{\hsize}{!}{
\includegraphics[width=\hsize,angle=270]{aceitunoFig10.ps}}
  \caption{\label{SBAM}
Distribution of the night sky surface brightness versus airmass
for two of the five filters installed in the instrument: the
B-band (top panel) and the V-band (bottom panel). The red
dashed-line shows the expected distribution in the absence of
light pollution, while the green solid-line shows the distribution
when a $\sim$15\% of light pollution is taken into account.
 }
  \end{figure}

\subsection{Instrumental zero-point}

Following the procedures described Section 2, once a frame is
obtained for a particular band, the reduction procedure corrects
it for the bias, the dark current and the flat-field. The
astrometric solution is applied to the frame, and, according to
the time and date, an automatic routine (similar to the one used
to derive the astrometric solution) cross-checks the expected
location of the stars in the photometric catalogue (Ducati 2002),
and looks for nearby bright stars. Finally, following the aperture
photometry technique described before, the program derives a final
catalogue with the positions, airmasses, apparent magnitudes of
the clearly identified stars, and the corresponding extinction at
the considered location.

A new zero point is derived automatically for every single
exposure, and the master value is updated if the atmospheric
conditions are the proper ones. To do so, the software estimates
the average extinction from the individual values derived at the
location of each star.  A clipping algorithm is applied to reject
those values different in more than a 10\% from the median value.
If more than a 75\% of the stars are within a 10\% of the median
value, the data are fitted adopting a linear regression based on
Equation 1. If the correlation coefficient of this regression is
larger than 0.95, the master zero point of the considered filter
is automatically updated. If not, the average extinction is
provided but it is not taken as a reference.

Figure 8 illustrates this process. In each panel of the figure it
is shown the distribution of the apparent magnitude versus the
airmass for four of the five filters installed in the instrument.
Only the stars following the previously described rejection
criteria are included in these figures. The solid line shows the
result of the linear regression process for the plotted data
points. Table 1 lists the values of the zero-points derived in the
first calibration run, at the Calar Alto observatory.

In addition to the zero-point, the calibration procedure provides
with an estimation of the global atmospheric extinction. However,
while the zero-point is quite stable, the extinction is a highly
variable parameter. Table 2 shows the comparison between the
extinction coefficients published by S\'anchez et al. (2007) for
the Calar Alto observatory, and those derived by ASTMON along the
Calibration Run in the night of the June 2nd, 2010, what was found
to be mostly photometric. Both estimations of the average
extinction are consistent one each other, showing that the
instrument provides reliable measurements of this parameter. The
simultaneous derivation of the extinction at different bands is
required to analyze the relative contribution of different
components to this extinction (scattering, dust attenuation and
ozone absorption), which we will analyze in forthcoming studies,
once we have collected an statistically significant number of
data.

%
%


\begin{table}
\begin{center}
\caption{Comparison of the Extinction Coefficients}
\label{tab_ext}
\begin{tabular}{ccc}\hline
\tableline\tableline
Band & CAHA  &  ASTMON  \\
     &  $\kappa_\lambda$  & $\kappa_\lambda$  \\
\tableline
B    & 0.27$\pm$0.10 & 0.25$\pm$0.03 \\
V    & 0.18$\pm$0.07 & 0.16$\pm$0.08 \\
I    & 0.08$\pm$0.05 & 0.11$\pm$0.01 \\
\tableline \tableline
\end{tabular}
\end{center}
\end{table}

Figure 9 shows the distribution of the night-sky surface brightness
derived by the instrument during the calibration run.  The red
dashed-line represents the expected night-sky surface brightness when
no light pollution is considered. This red dashed-line has been
derived assuming the airmass dependency of the natural night-sky
surface brightness described in Benn \& Ellison (1998a,b) and S\'anchez
et al. (2007,2008). The green solid line represents the expected
night-sky surface brightness assuming a contribution of the light
pollution of a 15\% of the total night-sky surface brightness at the
Zenith and assuming that it increases linearly with the airmass. We
still do not know how the light pollution increases with the airmass,
and therefore, this is the first attempt to model this
distribution. However, it is found that the measured values are capped
between both curves, and therefore, the contamination for the light
pollution should not be larger than the estimated value. This value is
of the order of the one reported by S\'anchez et al. (2007), on the
basis of the analysis of the night-sky spectrum at the observatory.


 \begin{figure}
\resizebox{\hsize}{!}
{\includegraphics[width=\hsize]{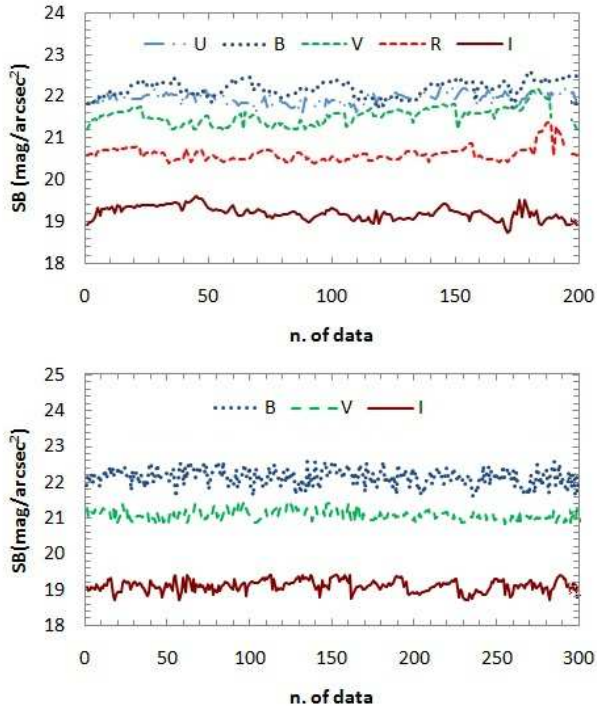}}

  \caption{\label{SBCahaDonana}
Monitoring of the night-sky surface brightness at the zenith for a
complete month. Upper panel: data obtained at the Calar Alto
observatory during January of 2011. Lower panel: data obtained at
the Do\~nana National Park during May of 2010.
 }
  \end{figure}

Many of the values reported on the night sky surface brightness
published so far are based in a few amount of values (eg.,
S\'anchez et al.  2007), and/or they have been taken with
different instruments using different techniques since they were
not the main goal of the observational projects (eg., Benn \&
Ellison et al. 1998a,b), which introduced errors difficult to
control. The instrument described in this article is able to
obtain a large number of estimations of the night sky surface
brightness along a single night, taken on purpose, and easy to
compare. With one acquisition per filter every minute, and a gap
of 300 seconds per filter set, the instrument is able to complete
a filter set (five filters) in $\sim$8 minutes. Therefore, it
provides $\sim$75 individual estimations of the night-sky surface
brightness per filter and night. With this amount of data it is
possible to trace variations of the night-sky brightness along
each night. Figure \ref{SBCahaDonana} shows an example of this
monitoring process. The figure shows a comparison of the variation
of the night-sky surface brightness at the zenit, monitored along
a complete month, for the two locations where we have currenly
installed an ASTMON unit, the Calar Alto Observatory and the
Do\~nana National Park. The data produced by long-term
monitorings, similar to the presented in this figure but expanded
along several years, will allow us to determine the differences in
the night-sky brightness and spectral energy distribution at
different locations, evaluating the seasonal and yearly evolution,
and their dependency with other parameters (like the solar
activity). We will have to acquire data for at least two years
before deriving a firm conclusion about the seasonal evolution of
the night-sky surface brightness, and for a period of more than
eleven years to determine the effects of the solar activity.

\begin{table}
\begin{center}
\caption{Comparison of the night-sky surface brightness}
\label{tab_mag}
\begin{tabular}{llll}\hline
\tableline\tableline

Band & ASTMON-Calar Alto $^*$ & ASTMON-Do\~nana $^{**}$ & CAHA$^{***}$ \\
\tableline
U & 22.02$\pm$0.19 &       -        & 21.81$\pm$0.10\\
B & 22.29$\pm$0.20 & 21.89$\pm$ 0.3  & 22.41$\pm$0.15\\
V & 21.54$\pm$0.10 & 20.95$\pm$0.15 & 21.53$\pm$0.18\\
R & 20.59$\pm$0.15 &       -        & 20.84$\pm$0.27\\
I & 19.11$\pm$0.23 & 19.09$\pm$0.25 & 18.70$\pm$0.85\\
\tableline
\end{tabular}
\end{center}

$*$ Data obtained at Calar Alto Observatory during all the
available dark time in January 2011.

$**$ Data obtained at the
Do\~nana National Park during all the available dark time in May
2010.

$***$ Data from the darkest night presented by S\'anchez et
al. (2007).

\end{table}

The night-sky surface brightness is one of the parameters that
qualifies better an astronomical site. In S\'anchez et al. (2007)
it was found that Calar Alto was among the best astronomical sites
in the world, regarding this parameter. To provide a fair
comparison, it is required to measure the night-sky brightness at
the zenith, corrected for extinction and in a dark night (eg; Benn
\& Ellison 1998a,b). Table 3 shows the comparison between the
values derived for the night-sky brightness using ASTMON during
all the available dark time in a complete month (seeing in Figure
10), and those reported in the literature (S\'anchez et al. 2007).
The values are almost identical for any of the considered bands.
The small discrepancies may be done due to the different seasons
when the data were collected.

{Figure 12 shows a typical example of the results obtained
by ASTMON during a normal working sequence. The skybrightness maps
were obtained in July 2010 at the Calar Alto night skies. The
lower row displays the raw images  while the upper one displays a
colorized sky brightness surface for the available sequence of
filter (B, V, and I are shown in the figure). Each color
represents different magnitudes per squared arcseconds and at a
first sight the brightness variances of the sky can be easily
identified.}

\begin{figure*}
\resizebox{\hsize}{!} {
\includegraphics[width=\hsize,angle=0]{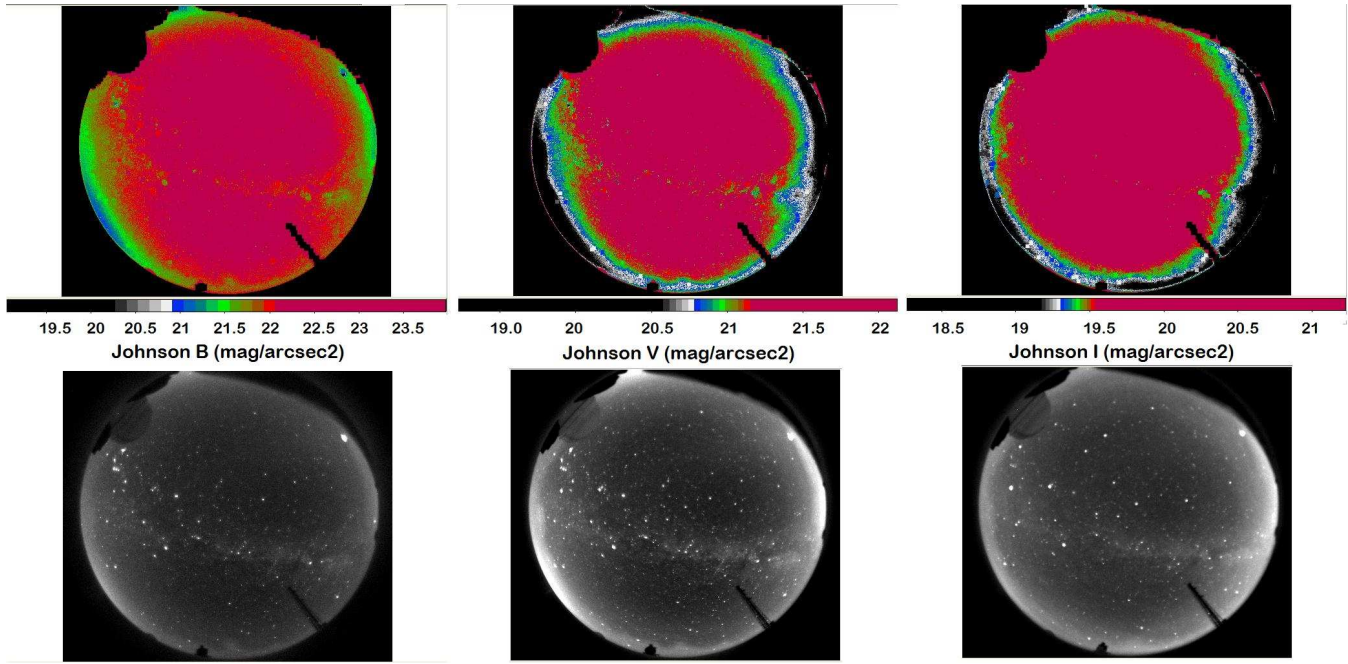}
}
 \caption{\label{sbmap}
The final result of ASTMON is to provide a completely
 processed sky brightness map sequence of the whole sky for all the available
 filters as it is shown in this figure. Upper panel: Processed
 sequence. Each color represents different skybrightness level.
 Low er panel: Raw images. Data obtained by ASTMON at the Calar
 Alto observatory during July of 2010.
 }
  \end{figure*}

{For a quantitative inspection of the mentioned variances,
the pixel's values of the FITS files contain the skybrightness
value expressed in the appropriate units (i.e. magnitudes per
squared arcseconds). This permits to display profiles along the
frame. Figure 13 shows a particular example about the influence of
the light pollution of a village being compared with the
skybrightness value at the zenith. This will allow to perform
detailed studies about the evolution in time, of the light
pollution produced by the surrounding areas where ASTMON is
located. A more detailed study of skybrightness in the areas
mentioned in this paper will be prepared in a near future.}

\begin{figure*}
\resizebox{\hsize}{!} {
\includegraphics[width=\hsize,angle=0]{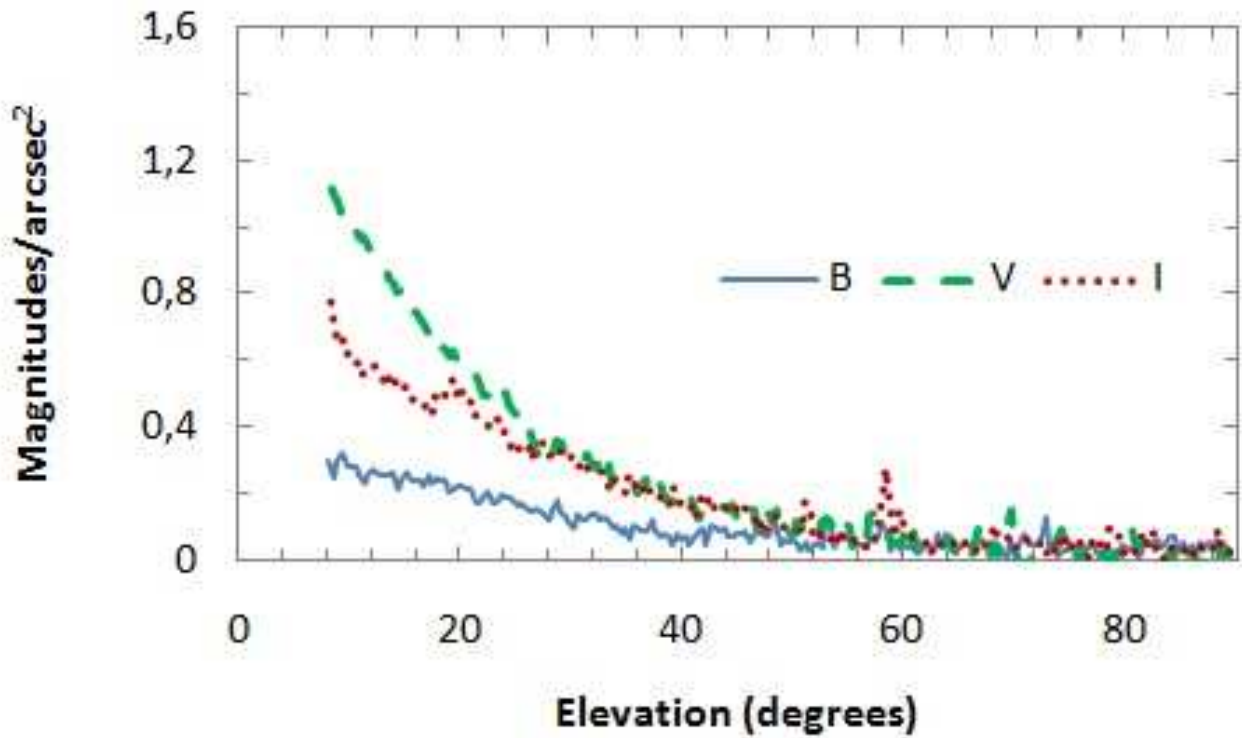}
}
 \caption{\label{matalascanas}
This figure shows the profile's variation of the zenith
skybrightness to the horizon level affected by the light pollution
of the near village Matalasca\~as in Huelva (Spain). Data obtained
in December of 2010 at the Do\~nana National Park (Huelva).
 }
  \end{figure*}

\section{Summary and Conclusions}
\label{sum}

We have developed and manufactured the first permanent station to
derive an exhaustive study of the light pollution and the
atmospheric extinction, by applying the standard techniques of
astronomical photometry. The system is based on a all-sky fisheye
lens, with a filter wheel, and a CCD detector that allows to
obtain automatic measurements of the night-sky surface brightness
in 5 different bands.  The system has been designed to afford
extreme weather conditions and it can be installed in a remote
place, being fully robotic. The instrument comprises a software
that it is fully automatic. It performs all the required tools to
operate the instrument, switching on and off automatically,
acquires the science frames in the corresponding filters, and
finally, reduces, calibrates and analyzes them to derive time
dependent maps of the night-sky surface brightness, and individual
estimations of atmospheric extinction for every band, together
with a rough estimation of the cloud coverage.

The early experiments performed during the calibration runs of the
instrument have demonstrated the validity of the conceptual design
and its implementation. The values provided by the instrument, for
both the night-sky surface brightness at the zenith, and the dust
extinction, are consistent with ones reported in previous
published results.

Currently, three ASTMON units have been installed at different locations in
Spain: (i) the Calar Alto Observatory (Almeria); (ii) the Do\~nana National
Park (Huelva) and (iii) the building of the Physics Department of the
Universidad Complutense de Madrid (Madrid). All of them are fully operational
and they are providing data for every clear night. A deep analysis of these
data will be presented in forthcoming articles, once we have acquired a
statistically significant number of them.

\section{Acknowledgments}

  JA and SFS thanks the sub-programs of {\it Viabilidad, Dise\~no, Acceso y
    Mejora de ICTS} , ICTS-2008-24 and ICTS-2009-32, the {\it PAI Proyecto
    de Excelencia} P08-FWM-04319 and the funds of the PAI research group
  FQM360, and the MICINN program AYA2010-22111-C03-03.

\newpage

 \label{lastpage}


\end{document}